# OSC Community Lab: The Integration Test Bed for O-RAN Software Community

Fransiscus Asisi Bimo*, Ferlinda Feliana*, Shu-Hua Liao*, Chih-Wei Lin*, David F. Kinsey†,
James Li‡, Rittwik Jana§, Richard Wright†, and Ray-Guang Cheng*,
* Department of Electronic and Computer Engineering, National Taiwan University of Science and Technology, Taiwan
† AT&T Inc., USA
‡ China Mobile Technology Inc., USA
§ Google Inc., USA

*Abstract*—O-RAN Software Community (OSC) is an open-source project collaborated by O-RAN Alliance and Linux Foundation, aiming to develop reference software components based on 3GPP and O-RAN Alliance specifications. The OSC has twelve projects. Among them, the Integration and Testing (INT) project is responsible for testing the requirements documented in each release for end-to-end and use case testing. Three OSC Community Laboratories were built to speed up the integration and interoperability testing among different projects. This paper summarizes the software components developed by OSC projects and the status of the three OSC Community Laboratories. The activities of each laboratory, how the community collaborates, and the challenges we encountered along the way were elaborated.

## I. INTRODUCTION

During the past few years, cellular has become a more flexible and programmable approach. A recent paradigm shift happened as Open RAN (O-RAN) movement started by O-RAN Alliance promotes openness by unlocking the proprietary interface between disaggregated RAN units, including the fronthaul between the Radio Unit and the Distributed Unit and midhaul between the Distributed Unit and the Centralized Unit. The concept of open interfaces enables vendor diversity in an initially-dominated RAN industry and interoperability to run on commercial-off-the-shelf (COTS) servers. At the same time, O-RAN also introduced intelligence features to ease-up controlling the increasingly complex RAN. Besides the open interface and intelligence, O-RAN emphasized open source software as one of its fundamental dimensions. [1]

Open source in telecom is enabled by the convergence between Information Technology and Telecom, as telecom technology shifts towards software [2]. Open source in cellular is expected to drive the advancement of cellular technology. The open source cellular technology will help the research community mitigate one of the challenges they face: limited accessibility to the cellular network. Two commonly known open source RAN software are SRSRAN and OpenAirInterface (OAI) [3].

O-RAN Alliance collaborated with The Linux Foundation to form O-RAN Software Community (OSC). There are two main purposes of this community. Firstly, OSC provides reference software that complies with the O-RAN specification. Developing the software allows direct feedback on the specification made by O-RAN Alliance. Secondly, the community aims to provide open source O-RAN that is publicly available. OSC software is developed by operators, vendors, and academia collaborate. Due to this, testing and integration of the whole parts of OSC O-RAN software architecture became a challenge as it needs to bond different components developed by different projects worked by different organizations/companies. [4]

Testing and integration of the whole parts of OSC O-RAN software architecture became a challenge as it needs to bond different components developed by different project developed by different organization/companies. A testbed is build to verify interoperability between components can run without any issues. It can also provide a common framework for discovering, reserving, using, and testing 5G networks and use cases. [5] The first testing laboratory of OSC is built at the AT&T facility in New Jersey. Two other facilities are also built at China Mobile in Silicon Valley and the National Taiwan University of Technology Lab in Taipei. The facilities are made to perform conformance testing and the interoperability test between different units.

This paper presents the approach OSC took to test and integrate each O-RAN component developed in the different project groups through the OSC labs. We present the overview of the components developed by each OSC project, as well as the challenges in performing testing and integration. In our paper, we hope to share our experience building the lab to test the software underlying the facilities of 5G testbeds, which is the RAN (O-RAN) itself.

The rest of the paper is organized as follows. In section II, we summarize the components developed by each OSC project. Section III presents the infrastructure of the three OSC Labs and the lessons we learned in establishing the Labs. These include interaction with stakeholders, technical cooperation among international developers, and technical lessons. Section IV summarizes our work and draws conclusions.

## II. OVERVIEW OF OSC PROJECTS

O-RAN Alliance collaborates with The Linux Foundation in forming OSC. Through this collaboration, OSC became a project under The Linux Foundation, and O-RAN Alliance acted as its sponsor [6]. This collaboration enables the O-RAN Software Community to use the legal framework and the project infrastructure of The Linux Foundation. In addition, the community also collaborates with some of the foundation's open-source projects, which include the Open Network Automation Platform (ONAP), Akraino, and Acumos AI where OSC contributes to these three projects.

OSC's mission is to develop open-source software that enables modular, open, intelligent, efficient, and agile radio access networks with alignment to the architecture specified by O-RAN Alliance. The OSC's leading committee, the Technical Oversight Committee (TOC), is appointed by O-RAN Alliance's Technical Steering Committee. TOC is responsible for all technical oversight of OSC. On behalf of OSC, the committee handles negotiation and prioritization with O-RAN Alliance. To plan the release of the software, TOC formed a subcommittee, the Requirements and Software Architecture Committee (RSAC). RSAC receives a recommendation from WG1 to include certain specifications or design features in a specific release, which they select based on the available resources and timelines of OSC. Pre-standards development and code contributions will be designed and specified by the projects in OSC and documented in the O-RAN SC Wiki. O-RAN Alliance receives the software release from OSC for testing the end-to-end use case. At the same time, they also receive contributions from non-OSC O-RAN members and specifications from 3GPP. The alliance will create O-RAN specifications, architecture, and reference design from these contributions, including the specification for testing to be defined by the Testing and Integration Focus Group (TIFG). The TIFG will request OSC's TOC to resolve standards or OSC variance in cases where the code diverges from the specification.

The first software version by OSC, 'Amber,' was released in late November 2019 with more than 1 million lines of code by over 60 developers from more than ten companies. There are currently five releases. The Linux Foundation Insight contributed around 4.02 million lines of code across 119 repositories from 240 contributors for the E Release. Thirteen projects are developing different parts of the specification, which division of contribution can be seen on each release and documented in the docs and wiki sub-domain of [7]. Two of the 13 projects in OSC are halted due to the lack of contributors. The 11 active projects comprised Near-Realtime RIC X-APPs (RICAPP), Near-Real-time RAN Intelligent Controller Platform (RICPLT), O-RAN Distributed Unit High Layers (O-DU HIGH), O-RAN Distributed Unit Low Layers (O-DU LOW), Operations and Maintenance (OAM), Simulations (SIM), Infrastructure (INF), Integration and Testing (INT), Documentation (DOC), Non-Realtime RAN Intelligence Controller (NONRTRIC), and Service Management and Orchestration (SMO). The key contributors for each project are AT&T (RICAPP), Nokia (RICPLT), Radisys (O-DU High), Intel (O-DU Low), highstreet technologies (OAM and SIM), Wind River (INF), China Mobile (INT and DOC), Ericsson (NONRTRIC), and VMware (SMO).

These projects realized the implementation of O-RAN architecture based on their respective scope. Figure 1 shows the relationship between OSC projects and O-RAN functional elements. The following subsections will provide further details on each OSC project.

On the management side, O-RAN architecture comprised the Service Management and Orchestration with Non-RT RIC residing as its part. **SMO** project handles the integration of different software artifacts of existing open-source projects. The SMO project aims to create a fully functional open-source SMO, supporting O1, O1/VES, O2, A1, and R1 interfaces. The development of the SMO project relates to WG1, WG2, WG5, and WG6. The project provided the containers for the O1 interface and VES collector. Since application on-boarding (rApps on non-RT RIC and xApps on near-RT RIC) is also the responsibility of SMO, they are also trying to define the application package scheme. SMO is collaborating with the OAM project to test and drive the published data models for the O-RAN solution.

The Operation and Maintenance (OAM) bridges the interworking between SMO and other O-RAN components. **OAM** project provides reference implementation according to the O-RAN OAM WG10 specification while also considering WG4 for O-DU Netconf configuration and TIFG. The project manages four repositories: the main repository of OAM, Network Function Adopter for model and protocol conversion (NF OAM Adopter), BBF-TR069 management interface converter to O-RAN OAM Interface specification (TR069 Adapter), and the repository for yang data models and yang-tools generated classes.

O-RAN Alliance defined a new intelligent unit called the RAN Intelligent Controller (RIC). RIC acts as software-defined networking (SDN) controller for O-RAN. Based on its behavior, RIC is divided into the Non-Realtime (Non-RT) RIC and the Near Realtime (Near-RT) RIC. Non-RT RIC resides in the SMO. The **NONRTRIC** project is responsible for implementing the functionalities of non-Realtime RIC based on WG2 specifications (including service and policy management, RAN analytics, and model training for the near-RealTime RICs). The scope of the project includes handling the functionality of the A1 interface that connects to Near-RT RIC and supporting rApps. This project focused on the use cases supported by ONAP. Non-RT RIC comprised the control panel/dashboard, R-App Catalog, A1 Enrichment Information (EI) Coordinator, two ONAP CSSDK/SDNC (A1 Policy Management Service and A1 Adapter), and A1 simulator, which simulates the A1 aspect of near-RT RIC.

SMO, OAM, Non-RT RIC, and simulators provided by the SIM project realized the closed-loop use case proposed by RSAC in D Release. The three are deployed as containers in Ubuntu.

The Near-RT RIC is a software platform that enables xApp to control O-DU and O-CU. The component is an independent entity, unlike the Non-RT RIC. Near-RT RIC software platform development is handled by the **RICPLT** project, while xApp development is handled by the **RICAPP** project. Both are deployed as Kubernetes clusters. RICPLT project focuses on developing functions according to WG3 specification and developing the O1 interface to connect with SMO, A1 interface to connect with non-RT RIC, and E2 interface to the O-DU and O-CU. RICAPP project is in charge of the open-source sample xApps and platform applications for integration, testing, and demonstration (mainly related to WG3 and use cases in WG1). Even though different projects develop it, both platform and xApp evolve each other. Since xApps utilizes the software platform's capabilities, typical tasks and code can be moved to the platform's SDK library for broader use (i.e., alarms and metrics).

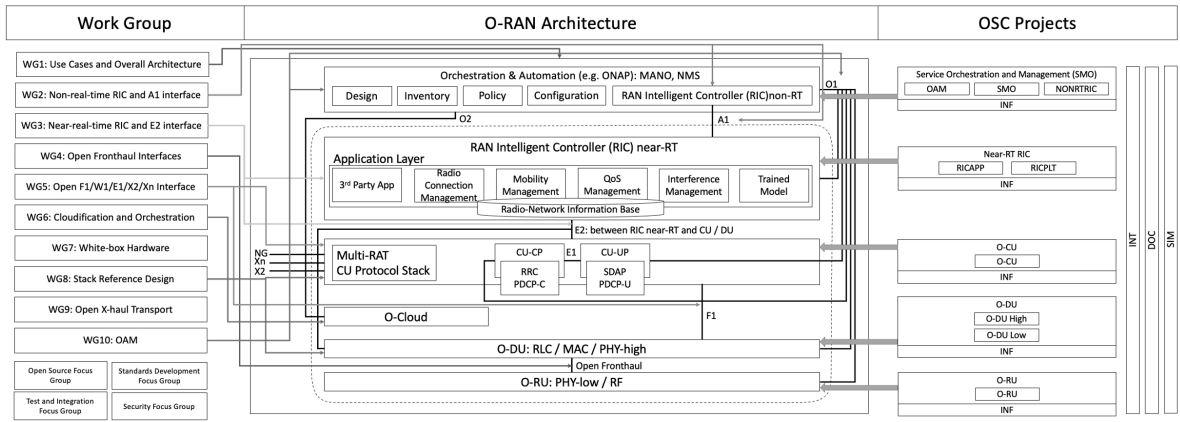

Fig. 1. Relationship between O-RAN Architecture and OSC Projects

There are several xApps currently available; the newest xApp (by E Release) is the RAN Control (RC) xApp by Mavenir. The RC xApp implements the subset E2SM-RC (E2 service model for RAN "Control") recently approved specification in Workgroup 3 [8]. The 11 xApps are contributed by six organizations (HCL, Samsung, AT&T, ChinaMobile, UTFPR, and Mavenir). xApps from different organizations can be combined as a use case. For instance, the anomaly detection use case combined Anomaly Detection (AD), QoE Predictor (QP), and Traffic Steering xApp. In the E release, several xApps are integrated with the Viavi simulator: Anomaly Detection (AD), QoE Predictor (QP), Traffic Steering (TS), Load Prediction (LP), RAN Control (RC), and KPI Monitor (KPIMON).

O-RAN 'RAN' components are adapted from the 3GPP-defined RU-CU-DU, which made up the basic architecture of a Next Generation Node B (gNB). Those components adopt O-RAN Alliance's specification along with functional split 7.2x, renamed 'O-RU,' 'O-DU,' and 'O-CU.' O-CU project and O-RU project in OSC is currently halted due to the lack of contributors, while the O-DU project has been active since Amber Release. The works for O-DU are separated into two development projects, as per the separation of layer 1 (l1) and layer 2 (l2) of O-RAN Alliance Specification: O-DU Low and O-DU High.

The **O-DU High** project is responsible for developing a protocol stack that complies with WG8 and the interfaces to communicate with other network functions. Radisys contributed the seed code for O-DU High in the first Amber Release. The source code for O-DU High is written in C. The project's initial releases focused on providing protocol stack implementation that complied with WG8, which includes the Radio Link Control (RLC), Medium Access Control (MAC), and the High Physical (High-PHY) layer. There are four interfaces the software currently supports with an ongoing enhancement: FAPI interface to O-DU Low by 5G FAPI Translator Module (TM), F1 interface to O-CU, E2 interface to the near-RT RIC, and O1 interface to the SMO. These interfaces followed specifications from WG3 and WG5 in addition to WG8. The software is comprised of multiple entities/modules that are separated into eight threads. The project goals are intelligent control and the ability to handle multiple use cases with different Quality of Service (QoS). This goal has implemented support for traffic steering use cases, closed-loop automation use case, network slicing, and a partial implementation of health check use cases. The implementation of O1 is contributed by HCL Technologies Ltd., which joined with Radisys to contribute to the O-DU High project since Cherry Release.

**O-DU Low** seed code is contributed by Intel, using their 4G and 5G baseband PHY Reference Design. It is in the form of a binary blob named l1app. The source code is written in C/C++. Implementation of O-DU Low only supports Intel Xeon® series Processor with Intel Architecture (either SkyLake or CascadeLake, minimum 2.0 GHz core frequency) running CentOS 7 (7.5+) [9]. The O-DU Low project is responsible for developing a protocol that complies with WG8 (l1 part) and the interfaces to communicate with other network functions. The software has implemented the relevant functions in 3GPP TS 38.211, 38.212, 38.213, 38.214, and 38.215. There are three interfaces O-DU Low supported with an ongoing enhancement: Front Haul interface to O-RU, FAPI interface to O-DU High, and interface to an accelerator (initially adopt DPDK BBDev). These interfaces followed WG4, WG8, and WG6, respectively. [7]

Testing and integration of OSC software are organized and supported by the **INT** project. In addition, they are also developing a tool for test flow definition and orchestration as a sub-project named Open Testing Framework (OTF), which is included as F Release focus. The testing and integration in OSC are divided into five levels:

1) Unit Testing: performed by developers by implementing test cases
2) Project level integrated testing: testing the internal flow of all software components that are deployed into a single system.
3) Project pairwise testing: testing conducted for the software of O-RAN components developing team to ensure communication compatibility.
4) System integrated testing: whole OSC software testing into a single system focusing on overall system health, inter-project communication/application compatibility, overall system deployment flow, and resource requirements.
5) Use case testing: testing of use case after system integrated testing is conducted by using the OTF.

Some of the tests of OSC software utilized the help of simulators. For example, is the A1 simulator used for functional testing (project-level) Non-RT RIC. **SIM** project is responsible for providing initial simulators used for testing O-RAN NF (Network Function) interfaces. The project is split according to the interfaces it simulates: A1, E1, E2, F1, FH, and O1. The structure of this project is flexible and changeable based on the needs of other projects.

System integration and use case testing are conducted using OSC Community testing lab [6]. Three OSC labs support the integration and testing of software developed by the community, two in the United States (New Jersey, Silicon Valley) and one in Taiwan (Taipei). Unlike OTIC, these labs are specifically built to support the development of OSC software.

III. OSC COMMUNITY LABS

OSC Community Labs played an important role in the development of the open-source reference implementation of the O-RAN solution. Every software should be able to interconnect with each other to accommodate the specification-defined use cases, and the community lab is where most software can be tested. The logical resources owned by each lab to be utilized for testing and integration are shown in Figure 3.

*A. OSC NJ Lab*

OSC's first testing and integration environment was built in T-Lab, a laboratory facility owned by AT&T. Over the years, AT&T's T-Lab has accumulated hardware such as RAN equipment by various vendors and 5G Core. The setup for this lab to be OSC's testing and integration environment began in September 2020, with O-DU being the first component to be installed in the lab by Intel (O-DU Low) and Radisys (O-DU High). The focus of OSC NJ Lab is the end-to-end (E2E) integration and testing of software components developed by the O-RAN Software Community. However, OSC software lacks the Core, O-CU, and O-RU parts to realize the E2E environment. In order to cover up for this, testing of the functionality of available software (O-DU) can be conducted where emulators and closed-source software will be used. The current O-CU residing in the OSC NJ Lab server is a closed-source/commercial O-CU from Radisys. Viavi Core Emulator covers the core network, O-RU uses Viavi TM500 O-RU Emulator, and the UE part uses Viavi TM500 UE Simulator. There is also a control PC for Viavi TM500. In the management and orchestration part, OSC provided a server for SMO, which includes the software artifacts made by SMO, OAM, and the Non-RT RIC project.

Figure 2 shows the architecture planned for the F release to provide an E2E integration and testing environment. The new addition planned to be added Front Haul Gateway (FHG) and Precision Time Protocol Grand Master (PTP-GM), which was delayed in E Release. The FHG has the capability of protocol translation, packet switching, and aggregating between multiple O-RUs to a single location consisting O-DUs. The PTP-GM is a device for X-haul timing and synchronization. OSC's O-DU will be connected to TM500 by FHG with synchronization between the two based on PTP-GM.

*B. Silicon Valley INT Lab*

The Silicon Valley INT Lab is located in Milpitas, California. The lab is built and maintained by China Mobile. Although it is still in working progress, the lab was established by mimicking the software and networking environments of the OSC community lab in Bedminster. At its initial stage as "software-centric," functional and performance testings on the OSC RIC platform and xApps are performed in the lab. Various OSC simulators (such as the E2Sim and O1 Simulator) are deployed to meet the end-to-end testing requirements for specific use cases. Figure 4 shows the current architecture of the lab.

In the near term, the lab's primary focus is on the SMO, Non-RT RIC, Near-RT RIC, and INF projects to validate/verify the installation/deployment of the module/services; and to carry out manual and automated close-loop use cases and integration testing. Since the O-Cloud support from the INF project is at the center of the OSC F-release, the lab will support the O-Cloud testing and help test other OSC components, such as RIC, work on O-Cloud. The stretch goal is to run the DU/CU/RIC Interoperability and usability test, as well as set up a CI/CD pipeline for continuous system tests in the lab with the help of partners.

*C. OSC Taiwan Lab*

OSC Taiwan Lab is located in Taipei, Taiwan. This lab is built under the coordination of the National Taiwan University of Science and Technology (NTUST) and National Yang Ming

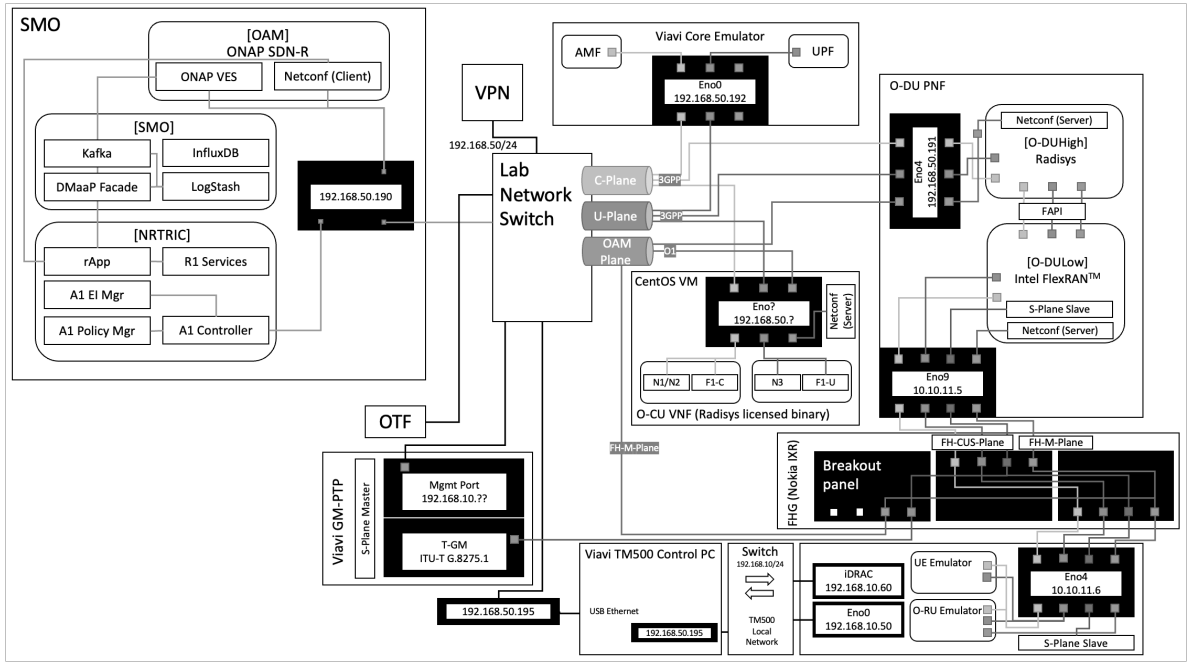

Fig. 2. OSC New Jersey Lab

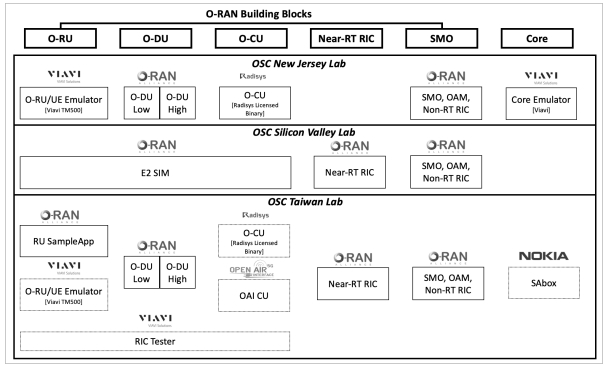

Fig. 3. Logical resources in each lab

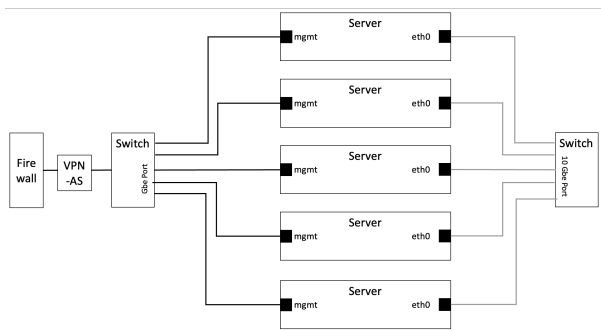

Fig. 4. Silicon Valley INT Lab

Chiao Tung (NYCU) and is supported by Chunghwa Telecom (CHT) and the 'B5G/6G Software Technology' project funded by the Ministry of Science and Technology, Taiwan. This lab focuses on development and integration testing. Similar to the Silicon Valley Lab, this lab aims to duplicate the environment in the New Jersey Lab but with a different focus on the integration part. This lab will focus on O-DU integration and security-related areas. In its setup phase, the lab has done several testing and integration for O-DU, including the TDD testing for O-DU in D Release and the limited end-to-end testing by utilizing simulators (RU SampleApp - O-DU - CU Stub).

In Figure 5 we show the architecture plan of the lab. On the top right corner is Nokia Bell Labs 5G SA Lite Core (SABOX), a commercial unit of core network from Nokia to support verification of backhaul connection to O-CU. The O-DU server is comprised of both O-DU High and O-DU Low. This server is currently connected to the RU SampleApp in another server using a direct connection. On the higher layer, O-DU High is connected to the near-RT RIC server through the E2 interface. By utilizing virtualization, the RIC server consists of Near-RT RIC platform and the Viavi RIC Tester, a commercial unit for testing RIC platform in order to develop both the RIC Platform and xApp. As timing synchronization is crucial in 5G RAN, this lab also has built-in PTP-GM inside FHG, which is connected to a GPS signal through an outdoor GPS/GNSS antenna.

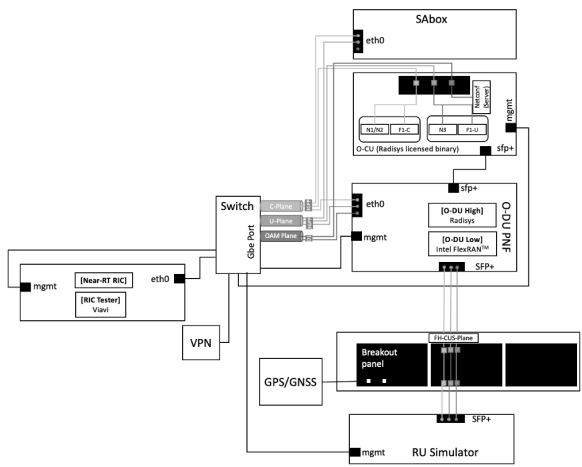

Fig. 5. OSC Taiwan Lab

*D. Lesson Learned*

*1) Interaction with stakeholders:* Aside from private companies, OSC Taiwan Lab has support from the Ministry of Science and Technology to build a 5G RAN environment to accelerate the development of O-RAN components. Apart from the technical achievement, we also shared our experience in the 5G Software Security Forum, supported by National Communication Commission, and discussed a topic related to 5G O-RAN software security on data communication. The talk is a part of awareness of possible security attacks on IoT, AI, and 5G technologies to build a safe, reliable, and resilient information and communication environment.

*2) International technical cooperation:* Integration and testing of software components developed by OSC are fundamental to ensuring that interworking between RAN components is proven well. The actual collaboration starts at this level where the three INT labs exist to provide space for integration done by each project to run several standard testing methods as a part of a development cycle. The collaboration also involves discussion in a bi-weekly meeting coordinated by OSC. In this meeting, every project leader shares the project status and milestones for upcoming releases and problem-solving in case of any issues.

*3) Technical lessons:* During the commissioning of the O-RAN components, several technical lessons were learned. For example, RAN components cannot run on all types of machines. There is a certain range of COTS server specifications that are supported. Integration and testing result helps to define the actual detailed specification of the supported COTS server. Other things worth noting are the exact version of the operating system's prerequisites, the proper configuration between connections, and accurate tuning parameters to compile the project. Without these practices could lead to failure setup. With its limitation, missing components of RU and CU left us the option to use the commercial product as a black box testing unit. There was a case of mismatch set up in VLAN ID, which led to a failure communication procedure when conducting end-to-end testing. To solve this issue, each project leader must disclose specific configurations from their testing to make integration successful.

## IV. CONCLUSION

In order to confirm system interoperability, each RAN project must test its component to verify the performance result and solve any integration issues. OSC Community Lab. testbeds is a substantial aspect of ensuring usability and interoperability between RAN components. OSC provided three labs to share the resources (servers, emulator hardware) for the community to support the development of OSC software. They accommodate developers to detect unexpected issues which occurs during testing and identify some limitation where these are essential things in software development. The three testbeds make focus integration and thorough evaluation on certain parts easier and enable efficiently allocating of resources. OSC New Jersey Lab is the most mature lab that has led the testing and integration of OSC releases. However, the lab itself is still under further enhancement to be full-fledged testing, integration, and demonstration platform, while the other two are still under development to follow suit. OSC Silicon Valley focused more on the intelligent aspect of O-RAN (OAM, SMO, Non-RT RIC), and OSC Taiwan Lab focuses on DU integration, testing, and security-related areas. Lastly, from the experiences, we want to encourage academia or industry to join and collaborate in developing O-RAN 5G architecture.